\documentclass[a4paper,11pt]{article}
\usepackage{pos}
\usepackage[backend=biber,url=false,doi=false,hyperref,style=numeric,sorting=none,bibencoding=utf8,maxnames=2,uniquename=init,giveninits=true]{biblatex}
\usepackage{tikz}
\usepackage{pgflibrarysnakes}
\usetikzlibrary{decorations.markings}
\usetikzlibrary{svg.path}
\usepackage{tikz-feynman}
\usepackage{amsmath}
\usepackage{bm}
\usepackage{simplewick}
\usepackage{booktabs}
\usetikzlibrary{patterns}
\usetikzlibrary{svg.path}
\usetikzlibrary{shapes,fit}
\usetikzlibrary{shapes.misc}

\AtEveryBibitem{\clearfield{title}}

\usetikzlibrary{fit}
\tikzset{%
  highlight/.style={rectangle,fill=gray,
    fill opacity=0.5,thick,inner sep=0pt}
}
\tikzset{cross/.style={cross out, draw=black, thick, minimum size=2*(#1-\pgflinewidth), inner sep=0pt, outer sep=0pt},
cross/.default={3pt}}

\tikzset{
    photon/.style={decorate, decoration={snake}, draw=black},
    electron/.style={draw=blue, postaction={decorate},
        decoration={markings,mark=at position .55 with {\arrow[draw=blue]{>}}}},
    gluon/.style={decorate, draw=magenta,
        decoration={coil,amplitude=4pt, segment length=5pt}}
}

\newcommand{\eval}[1]{\langle#1\rangle}

\newcommand{\Nf}{N_\mathrm{f}}

\newcommand{\tr}[1]{\operatorname{tr}\{#1\}}

\makeatletter
\DeclareFontFamily{OMX}{MnSymbolE}{}
\DeclareSymbolFont{MnLargeSymbols}{OMX}{MnSymbolE}{m}{n}
\SetSymbolFont{MnLargeSymbols}{bold}{OMX}{MnSymbolE}{b}{n}
\DeclareFontShape{OMX}{MnSymbolE}{m}{n}{
    <-6>  MnSymbolE5
   <6-7>  MnSymbolE6
   <7-8>  MnSymbolE7
   <8-9>  MnSymbolE8
   <9-10> MnSymbolE9
  <10-12> MnSymbolE10
  <12->   MnSymbolE12
}{}
\DeclareFontShape{OMX}{MnSymbolE}{b}{n}{
    <-6>  MnSymbolE-Bold5
   <6-7>  MnSymbolE-Bold6
   <7-8>  MnSymbolE-Bold7
   <8-9>  MnSymbolE-Bold8
   <9-10> MnSymbolE-Bold9
  <10-12> MnSymbolE-Bold10
  <12->   MnSymbolE-Bold12
}{}

\let\llangle\@undefined
\let\rrangle\@undefined
\DeclareMathDelimiter{\llangle}{\mathopen}%
                     {MnLargeSymbols}{'164}{MnLargeSymbols}{'164}
\DeclareMathDelimiter{\rrangle}{\mathclose}%
                     {MnLargeSymbols}{'171}{MnLargeSymbols}{'171}
\makeatother

\title{Efficiently unquenching QCD+QED at $\mathrm{O}(\alpha)$}
\author*[a]{Tim Harris}
\author[a]{Vera G\"ulpers}
\author[a]{Antonin Portelli}
\author[a,b]{James Richings}

\affiliation[a]{School of Physics and Astronomy, University of Edinburgh,\\
Edinburgh EH9 3FD, United Kingdom}
\affiliation[b]{EPCC, University of Edinburgh,\\EH8 9BT, Edinburgh, United Kingdom}

\emailAdd{tharris@ed.ac.uk}

\abstract{
    We outline a strategy to efficiently include the electromagnetic
    interactions of the sea quarks in QCD+QED.
    When computing iso-spin breaking corrections to hadronic quantities at
    leading order in the electromagnetic coupling, the sea-quark charges result
    in quark-line disconnected diagrams which are challenging to compute
    precisely.
    An analysis of the variance of stochastic estimators for the relevant
    traces of quark propagators helps us to improve the situation for certain
    flavour combinations and space-time decompositions.
    We present preliminary numerical results for the variances of the
    corresponding contributions using an ensemble of $\Nf=2+1$ domain-wall
    fermions generated by the RBC/UKQCD collaboration.
}

\FullConference{%
  The 39th International Symposium on Lattice Field Theory (Lattice2022),\\
  8-13 August, 2022 \\
  Bonn, Germany 
}

\begin{document}
\maketitle

\section{Introduction}
\label{sec:intro}

Several lattice QCD predictions which form important input for precision tests
of the Standard Model have uncertainties at or below the $1\%$ level, for
example the HVP contribution to $(g-2)_\mu$, $f_K/f_\pi$, $g_\mathrm{A}$ or the
Wilson flow scale $\sqrt{t_0}$ to name a
few~\cite{Aoyama:2020ynm,FlavourLatticeAveragingGroupFLAG:2021npn}.
However, to further improve such predictions, QCD with iso-spin symmetry is
not a sufficiently accurate effective description of the low-energy dynamics
and QED, which contributes one source of iso-spin breaking due to the different
up- and down-quark electric charges, must be included.
Recent efforts have been successful at including iso-spin breaking corrections,
and some of which fully account for the effects of the sea-quark electric
charges~\cite{Aoki:2012st,PhysRevLett.109.072002,CSSM:2019jmq,Bushnaq:2022aam,Borsanyi:2020mff}.
Nevertheless, many computations of iso-spin breaking effects still neglect to
incorporate these dynamical effects in an approximation known as
electroquenching.
As the FLAG report notes in Section
3.1.2~\cite{FlavourLatticeAveragingGroupFLAG:2021npn}, computations using the
electroquenched approximation might feature an uncontrolled systematic error.

In this work we aim to include the effects of the electric charge of the sea
quarks in the perturbative method known as the RM123 approach.
This amounts to computing at least two additional Wick contractions.
In order to sum the vertices in the resulting diagrams over the lattice volume,
some approximations must be used which often introduce additional fluctuations,
for example due to the auxiliary fields of a stochastic estimator.
Here we investigate some simple decompositions which may avoid large
contributions to the variance, so that sufficiently precise results can be
obtained to systematically include all sources of iso-spin breaking without
incurring a large computational cost.

\section{Sea-quark effects in the RM123 method}
\label{sec:rm123}

Due to the smallness of the fine-structure constant $\alpha\sim1/137$ and the
renormalized light-quark mass difference
$(m_\mathrm{u}^\mathrm{R}-m_\mathrm{d}^\mathrm{R})/\Lambda\sim 1\%$, it is
natural to expand physical observables (i.e. in QCD+QED) in these parameters to
compute iso-spin breaking corrections, as was first outlined in
Refs.~\cite{deDivitiis:2011eh,deDivitiis:2013xla}.
In the resulting expansion of an observable $O$
\begin{align}
    \eval{O} = \eval{O}\Big|_{e=0} + \tfrac{1}{2}e^2\Big[
        \frac{\partial}{\partial e}\frac{\partial}{\partial e}
        \eval{O}
    \Big]_{e=0} + \ldots
    \label{eq:rm123}
\end{align}
the leading corrections in the electric charge $e=\sqrt{4\pi\alpha}$ are
parameterized in terms of the correlation function
\begin{align}
    \frac{\partial}{\partial e}\frac{\partial}{\partial e}\eval{O} &=
    (-\mathrm i)^2\int\mathrm d^4x\,\int\mathrm
    d^4y\,\eval{J_\mu(x)A_\mu(x)J_\nu(y)A_\nu(y)O}_\mathrm{c}
    \label{eq:corr}
\end{align}
where the electromagnetic current for $\mathrm{u,d,s}$ quark flavours is defined
\begin{align}
    J_\mu &= \sum_{f=\mathrm{u,d,s}} Q_f \bar \psi_f\gamma_\mu\psi_f,\qquad
    Q_\mathrm{u}=\tfrac{2}{3},\quad Q_{\mathrm{d}}=Q_\mathrm{s}=-\tfrac{1}{3}.
    \label{eq:current}
\end{align}
By choosing the expansion point to be a theory with $\alpha=0$ and iso-spin
symmetry $m_\mathrm{u}=m_\mathrm{d}$, only correlation functions in the
$N_\mathrm{f}=2+1$ theory need to be evaluated, which we denote with
$e=0$ in Eq.~\eqref{eq:rm123}.
The precise definition of such a theory using an additional set of
renormalization conditions is necessary to fix the meaning of the leading-order
term on the right-hand side (and conversely the iso-spin breaking corrections
themselves).
Otherwise the predictions of QCD+QED are unambiguously defined, up to its
intrinsic accuracy, by fixing $\Nf$ quark masses and the QCD coupling as the
electric coupling does not renormalize at this order.
In the above, the ellipsis stands for the mass counterterms which are needed to
make physical predictions due to the contribution to the quark self-energy
induced by QED.

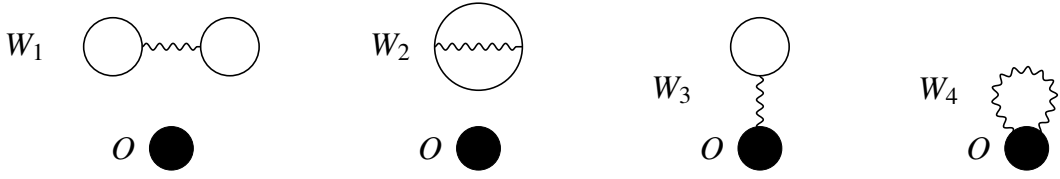
\begin{figure}[t]
    \centering
    \scalebox{0.5}{
    \begin{tikzpicture}
        \node (c) at (-6em,0) {\Huge $W_1$};
        \draw[very thick,->] (0,0) circle (2em);
        \draw[very thick,photon] (2em,0) -- (6em,0em);
        \draw[very thick,->] (8em,0) circle (2em);
        \draw[fill={black}] (4em,-7em) node[left=2em] {\Huge $O$} circle (1.5em);
%\node (a) at (-0em,-4em) {\Large $\ell - s$};
%\node (b) at (8em,-4em) {\Large $\ell - s$};
        \begin{scope}[xshift=25em]
            \node (c) at (-6em,0) {\Huge $W_2$};
            \draw[very thick,->] (0,0) circle (3em);
            \draw[very thick,photon] (-3em,0) -- (3em,0);
    %\node (a) at (-0em,-5em) {\Large $\ell,s$};
            \draw[fill={black}] (0,-7em) node[left=2em] {\Huge $O$} circle (1.5em);
        \end{scope}
\end{tikzpicture}}\hspace{4em}
\scalebox{0.5}{\begin{tikzpicture}
        \node (c) at (-6em,-3em) {\Huge $W_3$};
        \draw[very thick,->] (0,0) circle (2em);
        \draw[very thick,photon] (0,-2em) -- (0,-6em);
        \draw[fill={black}] (0,-7em) node[left=2em] {\Huge $O$} circle (1.5em);
%\node (a) at (4em,0) {\Large $\ell-s$};
\end{tikzpicture}}\hspace{4em}
\scalebox{0.5}{\begin{tikzpicture}
        \node (c) at (-6em,+4em) {\Huge $W_4$};
        \draw[very thick,photon] (0,0) .. controls (-7em,7em) and (7em,7em) ..  (0,0);
        \draw[fill={black}] (0,0) node[left=2em] {\Huge $O$} circle (1.5em);
\end{tikzpicture}}
    \caption{Wick contractions which appear at leading order in the expansion
        of a hadronic observable $O$ in the electromagnetic coupling.
        Each closed fermion line has contributions from all of the quark flavours
    $\mathrm{u,d,s,\ldots}$ with the appropriate charge factors.}
    \label{fig:wick}
\end{figure}

After integrating out the fermion and photon fields, the resulting Wick
contractions $W_i$ are shown in Fig.~\ref{fig:wick}, which contribute to the
derivative with respect to the electric charge through the connected
correlation function
\begin{align}
    \frac{\partial}{\partial e}\frac{\partial}{\partial e}\eval{O} &=
    \sum_{i=1}^4 \eval{OW_i}_\mathrm{c}.
    \label{eq:eobs}
\end{align}
The first two subdiagrams, which arise soley from the electric charges of the
sea quarks, can be expressed in terms of a convolution with the photon
propagator (in some fixed gauge) $G_{\mu\nu}(x)=\eval{A_\mu(x)A_\nu(0)}$
\begin{align}
    W_{1,2} = -a^8 \sum_{x,y}H^{\mu\nu}_{1,2}(x,y) {G_{\mu\nu}(x-y)},
\end{align}
where  $H_{1,2}$ are the traces of quark propagators
$S_f(x,y)=\eval{{\psi}{_f(x)}{\bar\psi}_f(y)}$
\begin{align}
    H_1^{\mu\nu}(x,y) &= \sum_{f,g}Q_fQ_g\tr {\gamma_\mu S_f(x,x)}\tr {\gamma_\nu S_g(y,y)},\\
    H_2^{\mu\nu}(x,y) &= -\sum_fQ_f^2\tr {\gamma_\mu S_f(x,y) \gamma_\nu
    S_f(y,x)}.
\end{align}
These two diagrams are the main subject of these proceedings, and the
techniques advocated for the first can be effectively reused for the third
diagram, $W_3$.
In the following sections we introduce stochastic estimators only for the quark
lines and compute the subdiagrams by convoluting with the exact photon
propagator which avoids introducing additional stochastic fields for the
$\mathrm U(1)$ gauge potential.
The final diagram $W_4$, which only contributes if the observable $O$ depends
explicitly on the (charged) fermion fields, is the only one surviving the
electroquenched approximation, and, can in most cases be computed efficiently
provided that the leading-order diagram is already under control.

We note that the variance of the contributions to the connected correlation
functions on the r.h.s. of Eq.~\eqref{eq:eobs} crudely factorizes
\begin{align}
    \sigma^2_{OW_{1,2}}&\approx
    \eval{O^2}_\mathrm{c}\eval{W_{1,2}^2}_\mathrm{c} + \ldots\\ %\eval{OW_{1,2}}_\mathrm{c}^2\\
    &\approx \sigma^2_{O}\sigma^2_{W_{1,2}},
    \label{eq:fact}
\end{align}
where in the first line we have made the Gaussian approximation, and in
the second line we have assumed that the fluctuations are much larger than the
signal $\eval{OW_{1,2}}_\mathrm{c}$.
Thus, in the following sections we will analyse the variance of individual
subdiagrams $W_{1,2}$ in order to gain a rough insight into the fluctuations of
the total correction, in a similar fashion to the analysis of
Ref.~\cite{Giusti:2019kff}.
In that case, however, the correction to the factorization of the variance is
exponentially suppressed in the separation between the vertices of the
subdiagrams.

\section{Quark-line disconnected subdiagram \texorpdfstring{$W_1$}{W1}}
\label{sec:w1}

We begin by noting that the hadronic part of the diagram factorizes into two
traces,
\begin{align}
    H^{\mu\nu}_1(x,y) &= T_\mu(x)T_\nu(y),
\end{align}%
each of which, with the current defined in Eq.~\eqref{eq:current} and in the
$\Nf=2+1$ theory with iso-spin symmetry, is the difference of the light- and
strange-quark propagators
\begin{align}
    T_\mu(x) &= \tfrac{1}{3}\tr{\gamma_\mu [S_\mathrm{ud}(x,x)-S_\mathrm{s}(x,x)]}.
\end{align}%
It is convenient to rewrite this difference as a product~\cite{Giusti:2019kff}
\begin{align}
    S_\mathrm{ud}-S_\mathrm{s} &= (m_\mathrm{s}-m_\mathrm{ud})S_\mathrm{ud}S_\mathrm{s}
    \label{eq:diffprod}
\end{align}
which makes the explicit suppression of $T_\mu$ in the
$\mathrm{SU}(3)$-symmetry breaking parameter $m_\mathrm{s}-m_\mathrm{ud}$
explicit.
This additionally results in a suppression of the variance of $W_1$ by
$(m_\mathrm{s}-m_\mathrm{ud})^4$.
This suppression results in a cancellation of a quartic short-distance
divergence in the variance of the contribution of each individual flavour to
$W_1$, explaining this favourable flavour combination.

While the identity in Eq.~\eqref{eq:diffprod} is easily derived for Wilson-type
fermions, here we sketch that it holds exactly for the domain-wall fermion
valence propagator $S_f={\tilde D_f}^{-1}$ which (approximately) satisfies the
Ginsparg-Wilson relation~\cite{Capitani:2000xi}.
Recalling the definition of $\tilde D_f$ in terms of the $5\mathrm D$ Wilson
matrix $D_{5,f}$ (see Ref.~\cite{RBC:2014ntl} for unexplained notation)
\begin{align}
    \tilde D^{-1}_f = (\mathcal P^{-1}D_{5,f}^{-1}R_5\mathcal P)_{11},
\end{align}
where the matrix indices indicate the coordinate in the fifth dimension, the
result is obtained immediately from
\begin{align}
    \tilde D_\mathrm{ud}^{-1} - \tilde D_\mathrm{s}^{-1} &= (m_\mathrm{s}-m_\mathrm{ud})
    (\mathcal P D_{5,\mathrm{ud}}^{-1}R_5D_{5,\mathrm{s}}^{-1}R_5)_{11}
\end{align}
by noting that the following matrix projects on the physical boundary
\begin{align}
    (R_5)_{\cdot\cdot} = (R_5\mathcal P)_{\cdot 1}(\mathcal P^{-1})_{1\cdot}.
\end{align}
The preceding identity is easily demonstrated using the explicit representations
\begin{align}
    R_5 = 
    \begin{pmatrix}
            &  &  & P^+ \\
            &  &  &     \\
        P^- &  &  &     \\
    \end{pmatrix},
    \qquad
    \mathcal P^{-1} = 
    \begin{pmatrix}
        P_- &        &        & P_+ \\
        P_+ & \ddots &        &     \\
            & \ddots &        &     \\
            &        & P_+    & P_- \\
    \end{pmatrix},
\end{align}
where $P_\pm = 1\pm\gamma_5$.

Using the identity for the difference, there are two independent estimators for the trace
\begin{align}
    \label{eq:standard}
    \Theta_\mu(x) &= \tfrac{1}{3}(m_\mathrm{s}-m_\mathrm{ud})\frac{1}{N_\mathrm{s}}
        \sum_{i=1}^{N_\mathrm{s}} \eta_i^\dagger(x) \gamma_\mu \{S_\mathrm{ud}S_\mathrm{s}\eta_i\}(x), \\
    \mathcal T_\mu(x) &= \tfrac{1}{3}(m_\mathrm{s}-m_\mathrm{ud})\frac{1}{N_\mathrm{s}}
    \sum_{i=1}^{N_\mathrm{s}} \{\eta_i^\dagger S_\mathrm{s}\}(x) \gamma_\mu \{S_\mathrm{ud}\eta_i\}(x),
    \label{eq:spliteven}
\end{align}
where the auxiliary quark fields $\eta_i(x)$ have zero mean and finite
variance.
The properties of both estimators were investigated in detail in
Ref.~\cite{Giusti:2019kff}, where it was shown that the contribution to the
variance from the auxiliary fields for the second split-even estimator was in
the region of a factor $\mathrm{O}(100)$ smaller than the first standard
estimator, which translates into the same factor reduction in the cost.
The split-even estimator has since been used extensively for disconnected
current correlators~\cite{SanJose:2022yad,Chao:2021tvp,Salg:2022poa}, while in
the context of the twisted-mass Wilson formulation similar one-end trick
estimators have often been employed for differences of twisted-mass
propagators~\cite{ETM:2008zte}.

In this work we propose an estimator for the first diagram $W_1$ using
\begin{align}
    \label{eq:w1}
    \mathcal W_1 &\approx 
    \Big(a^4\sum_x \mathcal T_\mu(x)\Big)
    \Big(a^4\sum_y \mathcal T_{\nu}(y)G_{\mu\nu}(x-y)\Big)
\end{align}
where independent estimators are used for the two traces to avoid incurring a
bias with a finite sample size.
The convolution in the second parentheses can be efficiently computed using the
Fast Fourier Transform (FFT).
With a minor modification, an estimator using all possible unbiased
combinations of samples can be written at the cost of performing $\mathrm
O(N_\mathrm{s})$ FFTs.
The standard estimator is obtained by replacing both occurances of $\mathcal
T_\mu$ with $\Theta_\mu$ in Eq.~\eqref{eq:w1}.

\begin{table}[t]
    \centering
    \begin{tabular}{cccccc}
        \toprule
        $L/a$ & $T/a$   & $m_\pi$             & $m_\pi L$   & $a$                & $N_\mathrm{cfg}$\\
        \midrule
        $24$  & $64$    & $340~\mathrm{MeV}$ & $4.9$       & $0.12~\mathrm{fm}$  & $50$ \\
        \bottomrule
    \end{tabular}
    \caption{The parameters of the C1 ensemble of $\Nf=2+1$ Shamir
        domain-wall fermions used in the numerical experiments in this
        work, see Ref.~\cite{RBC-UKQCD:2008mhs} for details.
    }
    \label{tab:ens}
\end{table}

\begin{figure}[h]
    \centering
    \includegraphics[scale=0.85]{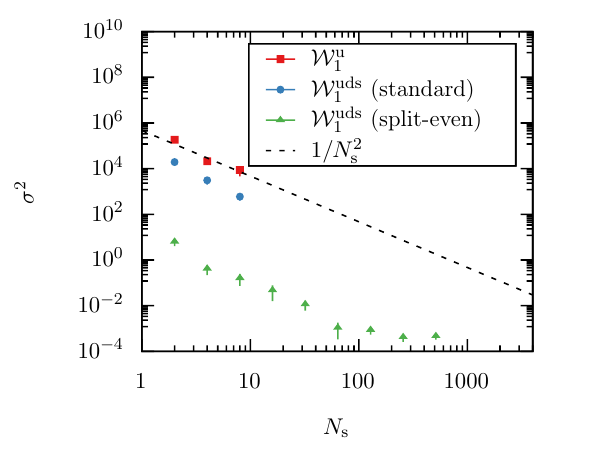}%
    \caption{Left: Comparison of the variance versus the number of sources
        for the $W_1$ quark-line disconnected diagram, using a single flavour
        (red squares), the standard estimator for $\mathrm{u,d,s}$ flavours
        (blue circles) and the split-even estimator (green triangles).
        The dashed line shows $1/N_\mathrm{s}^2$ scaling.
        In this figure, the (local) currents are not renormalized and the
        charge factors are not included.
    }
    \label{fig:specs}
\end{figure}

We performed an analysis of the variance for the standard and split-even
estimators for $\mathcal W_1$ using the domain-wall ensemble generated by the
RBC/UKQCD collaboration whose parameters are listed in Tab.~\ref{tab:ens}.
The photon propagator is computed in the $\mathrm{QED}_L$
formulation~\cite{Hayakawa:2008an} in the Feynman gauge.
The results for the variances, which are dimensionless numbers, are shown in
Fig.~\ref{fig:specs}.
In addition, we plot the variance for the contribution of a single flavour
$\mathcal W^\mathrm{u}_1$ using the standard estimators for the traces.
We note that all the variances are dominated by the fluctuations of the
auxiliary fields for small $N_\mathrm{s}$, and in particular scale like
$1/N_\mathrm{s}^2$ in that region.

As expected, the standard estimator including the light-quark and strange-quark
contributions (blue circles) is suppressed with respect to the
contribution of a single flavour (red squares).
Furthermore, the variance of the split-even estimator (green triangles) is
reduced by a factor of $10^4$ with respect to the standard one (blue circles).
This reduction is commensurate with the reduction in the variance observed for
the disconnected contribution to the current correlator~\cite{Giusti:2019kff},
which suggests the same mechanisms are present here.
For $N_\mathrm{s}\sim 100$, the variance is independent of the number of
auxiliary field samples which indicates that it is dominated by the
fluctuations of the gauge field.
In this case no further variance reduction is possible for a fixed number of
gauge configurations.
Finally we note that the convolution of the second parentheses of
Eq.~\eqref{eq:w1} can be simply inserted sequentially in any of the diagrams of
type $W_3$.

\section{Quark-line connected subdiagram \texorpdfstring{$W_2$}{W2}}
\label{sec:w2}

In contrast to the quark-line disconnected subdiagram, there is no cancellation
in the variance in the connected subdiagram $W_2$ between the light and
strange-quark contributions.
In this case, power counting suggests that the variance diverges with the
lattice spacing like $a^{-4}$ as $a\rightarrow 0$ and is expected to be
dominated by short-distance contributions.
Translation averaging should therefore be very effective and one way to
implement it is to use an all-to-all estimator~\cite{deDivitiis:1996qx} for the
quark propagator
\begin{align}
    \mathcal S_f(x,x+r) &= 
    \frac{1}{N_\mathrm{s}}\sum_{i=1}^{N_\mathrm{s}}\{S_f\eta_i\}(x)\eta^\dagger_i(x+r),
\end{align}
using independent fields for each propagator in the trace
\begin{align}
    {\mathcal H^{\mu\nu}_2(r)} &= a^4\sum_x \sum_f Q_f^2
    \mathrm{tr}\{\gamma_\mu
    \mathcal S_f(x,x+r)\gamma_\nu \mathcal S_f(x+r,x)\}.
\end{align}
As written, the estimator is feasible to compute for a small number of
separations $r$ between the vertices and, although it introduces a (mild)
signal-to-noise ratio problem at large $r$, should be efficient at small
$|r|\leq R$ given the leading extra contribution vanishes like
$N_\mathrm{s}^{-2}$, c.f.  Sec.~\ref{sec:w1}.

For the remainder $|r|>R$, we propose using $N_X$ randomly selected point
sources $X_n$~\cite{Li:2020hbj}
\begin{align}
    \bar H_2^{\mu\nu}(r) = \frac{L^3T}{N_X}\sum_{n=1}^{N_X}H_2^{\mu\nu}(X_n,X_n+r)
\end{align}
so that the total is split between short- and long-distance contributions
\begin{align}
    \label{eq:burger}
    \mathcal W_2 = a^4\sum_{|r|\leq R}{\mathcal H_2(r)}G_{\mu\nu}(r)
    + a^4 \sum_{r> R} \bar H_2^{\mu\nu}(r)G_{\mu\nu}(r),
\end{align}
using the efficient stochastic estimator for the noisy short-distance
contribution.
Ref.~\cite{Blum:2019ugy} introduced an importance sampling based on current
separations for higher-point correlation functions, whereas in this case we
make the separation based on the expected contributions to the variance. 
This approach avoids completely factorizing the trace which would require
either $\mathrm O(V)$ contractions or $\mathrm O(N_\mathrm{s}^2)$ FFTs to
include the photon line which we deemed unfeasible.

\begin{figure}[t]
    \centering
    \includegraphics[scale=0.75]{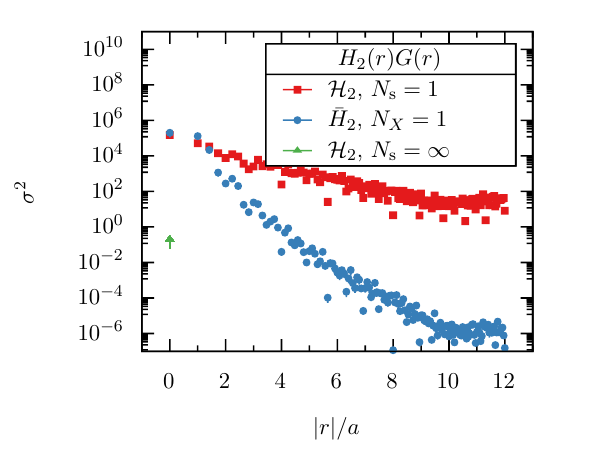}%
    \includegraphics[scale=0.75]{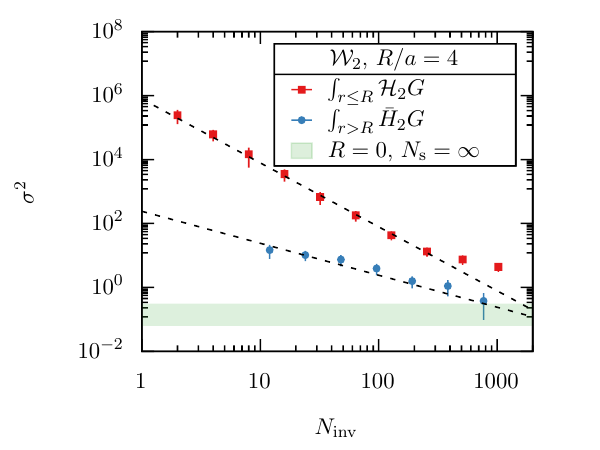}
    \caption{Left: the variance for the stochastic estimator (red squares) and
        point source estimator (blue circles) for the minimum number of
        inversions required, for the contribution with fixed separation between
        the currents $|r|$.
        The green triangle indicates the gauge variance for the point $r=0$.
        Right: the variance for the short-distance (red squares) and
        long-distance (blue circles) for the choice $R/a=4$, versus the number
        of inversions.
        The green band indicates the gauge variance for the contribution from
        $r=0$ only.
        The dashed lines indicate the expected leading $N_\mathrm{inv}^{-2}$
        and $N_\mathrm{inv}^{-1}$ scaling for the short- and long-distance
        components.
    }
    \label{fig:burger}
\end{figure}

In Fig.~\ref{fig:burger} (left) we illustrate the variance of each of the terms
in Eq.~\eqref{eq:burger} for the sum over a fixed separation $|r|$ between the
currents, for the case $N_\mathrm{s}=N_{X}=1$.
As expected, the variance from the contribution around $|r|\sim0$ dominates
both the stochastic (red squares) and point source estimator (blue circles),
and we observe the mild signal-to-noise ratio problem in the stochastic
estimator.
The green triangle denotes the gauge variance for the case $r=0$, which is
approximately suppressed by $(L^3T)/a^4$ compared to $N_X=1$ indicating
translation averaging is very effective for the short-distance contribution.
In the right-hand panel, we see variance of the short- and long-distance
contributions with the choice $R/a=4$ as a function of the number of inversions
(where $N_{X}=1$ corresponds to 12 inversions).
The variance is dominated by the short-distance contribution (red squares)
which however scales favourably like $N_\mathrm{inv}^{-2}$, while the
long-distance contribution (blue circles) which scales only like
$N_\mathrm{inv}^{-1}$ is much suppressed.
Deviations from the former scaling indicate that the gauge variance may be
reached with just $N_\mathrm{inv}\sim 1000$, which although is larger than
required for $W_1$ is still achievable with modern computational resources, and
universal for all observables.

\section{Conclusions}
\label{sec:conc}
In this work we have examined the Wick contractions which arise due to the
charge of the sea quarks in the RM123 method.
Such diagrams contribute, in principle, even to observables constructed from
neutral fields and are therefore ubiquitous in the computation of iso-spin
breaking corrections.
We have proposed stochastic estimators for the quark lines in such diagrams
which completely avoids the need to sample the Maxwell action stochastically,
thus eliminating one additional source of variance.
As for the case of disconnected contributions to current correlators, we have
shown it is beneficial to consider certain flavour combinations which have
greatly suppressed fluctuations.
We have shown that the split-even estimators generalize also to domain-wall
fermions and perform well compared with na\"ive estimators.
Thus the frequency-splitting strategy of Ref.~\cite{Giusti:2019kff} should
generalize appropriately for this fermion formulation.
In the second topology, however, there is no cancellation of the short-distance
effects in the variance by considering multiple flavours.
In this case, we propose decomposing the diagram into a short-distance part to
be estimated stochastically and a long-distance part estimated using
position-space sampling.
The variance is reduced sufficiently so that the gauge variance can be reached
with a reasonable computational cost.
Given their short-distance nature, these estimators should also succeed with
smaller quark masses, and furthermore as the diagrams are universal to all
iso-spin breaking corrections we anticipate that these simple decompositions
ought to be beneficial in large-scale simulations.
In particular we are developing these methods for refinements of our
computations of iso-spin breaking corrections within the RBC/UKQCD
collaboration, for example to meson (leptonic) decay
rates~\cite{Boyle:2022lsi,Boyle:2022sgz}.

\paragraph{Acknowledgments}
We use the open-source and free software Grid as the data parallel C++ library
for the lattice computations~\cite{Boyle:2016lbp}.
The authors warmly thank members of the RBC/UKQCD collaboration for valuable
discussions.
T.H., A.P. and V.G. are supported in part by UK STFC 1039 grant ST/P000630/1.
A.P. and V.G. received funding from the European Research Council (ERC) under
the European Union’s Horizon 2020 research and innovation programme under grant
agreement No 757646 and A.P. additionally under grant agreement No 813942.
This work used the DiRAC Extreme Scaling service at the University of
Edinburgh, operated by the Edinburgh Parallel Computing Centre on behalf of the
STFC DiRAC HPC Facility (www.dirac.ac.uk).
This equipment was funded by BEIS capital funding via STFC capital grant
ST/R00238X/1 and STFC DiRAC Operations grant ST/R001006/1. DiRAC is part of the
National e-Infrastructure.

\printbibliography

\end{document}